\documentclass[conference]{IEEEtran}
\IEEEoverridecommandlockouts

\usepackage{amsmath,amsfonts,amssymb}
\usepackage{amsthm}
\usepackage{array}
\usepackage{graphicx}
\usepackage{algorithm}
\usepackage{algpseudocode}
\usepackage{booktabs}
\usepackage{textcomp}
\usepackage{url}
\usepackage{cite}

\newtheorem{theorem}{Theorem}
\newtheorem{corollary}{Corollary}

\begin{document}
\flushbottom

\title{Joint Trajectory, Sensing, and Computation  \\ Co-Design for UAV-Assisted Federated Edge Vision}

\author{%
\IEEEauthorblockN{Kaichen Liu, Yueling Che, Sheng Luo}
\IEEEauthorblockA{\textit{College of Computer Science and Software Engineering} \\
\textit{Shenzhen University}, Shenzhen, China \\
2400101033@mails.szu.edu.cn; \{yuelingche, sluo\}@szu.edu.cn}
\and
\IEEEauthorblockN{Jianhua Tang}
\IEEEauthorblockA{\textit{Shenzhen Smart City Technology Development} \\
\textit{Group Company, Ltd.}, Shenzhen, China \\
jtang4@e.ntu.edu.sg}}

\maketitle

\begin{abstract}
This paper investigates UAV-assisted federated edge vision for training computationally heavy object detection models. Existing studies treat training data as static, ignoring the tight coupling between dynamic visual sensing and massive in-place computational burdens. To minimize the expected training time under battery constraints, we propose a four-phase joint trajectory, sensing, and computation co-design. To mitigate data heterogeneity, we use  a uniform sensing probability to translate federated convergence requirements into UAV mobility constraints. Furthermore, a two-layer alternating optimization (AO) algorithm is developed to adaptively determine the local data workload, optimally balancing the required training rounds against in-place computation delays.
Simulations validate the  superiority of our approach in model  convergence and accuracy over  baselines.

\end{abstract}


\section{Introduction}\label{sec:intro}

\IEEEPARstart{U}{nmanned} aerial vehicles (UAVs) equipped with high-resolution cameras are emerging as ubiquitous platforms for industrial infrastructure inspection \cite{jiang2021road}. This naturally drives the evolution of UAV-assisted federated edge learning (FEEL) toward federated edge vision, a paradigm that collaboratively trains computationally heavy object detection models (e.g., YOLO11x) across distributed UAVs without sharing raw images \cite{dianxing}. However, unlike conventional federated learning with static datasets, the data acquisition in mobile vision is an active sensing process. The UAVs' trajectories directly dictate the visual sensing quality and data volumes, which in turn determine the local computation workload and the ultimate learning convergence. Hence, isolated optimizations are inherently suboptimal, necessitating a rigorous joint trajectory, sensing, and computation co-design to minimize the total training time under stringent battery constraints.

Existing studies on UAV-assisted FEEL largely take the training data as given. They primarily focus on embedding FEEL into UAV networks for communication efficiency and privacy (e.g.,  \cite{zaoqi1,zaoqi2}), or integrating UAV trajectory and resource scheduling into the FEEL process to accelerate model convergence (e.g., \cite{22,24,wang2025uav}). To explicitly model the data acquisition phase, recent advancements have introduced Integrated Sensing, Communication, and Computation (ISCC) to the FL paradigm. For instance, the authors in \cite{tang2025integrated} and \cite{zhang2026integrated} studied UAV-assisted  ISCC for FEEL  that jointly optimizes sensing and communication resources; however, \cite{tang2025integrated} and \cite{zhang2026integrated} fundamentally considered a static UAV performing simple radar sensing. When applied to mobile visual sensing, these approaches fall short.
As noted in  \cite{drones9090595} and \cite{tongxing}, visual data quality and volume are inherently bounded by the UAVs' flight trajectories, since their altitude and speed govern the camera footprint, ground sampling distance (GSD), and statistical data heterogeneity.
To the best of our knowledge, existing ISCC frameworks have yet to accurately model this unique trajectory-sensing coupling in mobile vision, nor have they accounted for the massive in-place computational burden induced by modern deep vision models. As a result, developing a rigorous co-design that tightly couples the UAVs' mobility with sensing, computation, and communication remains a critical open challenge.

In this paper, we propose a joint trajectory, sensing, and computation co-design framework to minimize the expected training time in UAV-assisted federated edge vision. We propose a tractable four-phase operational model that  quantifies the time and energy coupling across communication, dynamic visual sensing, and in-place computation. To mitigate sensing data heterogeneity, we enforce a uniform sensing probability across all UAVs, which translates the federated convergence requirements into strict UAV mobility constraints. We formulate the training time minimization problem as  a mixed-integer nonconvex problem and develop a two-layer alternating optimization (AO) algorithm to solve it efficiently. In particular, our algorithm adaptively bounds the local data workload to optimally trade off the required training rounds against the heavy in-place computation delay.
Extensive simulations demonstrate that our co-design substantially improves model convergence accuracy over  baselines.

\section{System Model}
\label{sec:model}

\begin{figure}[!t]
    \centering
    \includegraphics[width=0.86\columnwidth]{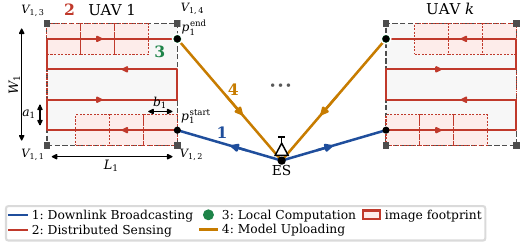}
    \caption{Illustration of the four phases in a training round, depicted via the trajectory projection of each UAV.}
    \label{fig:model}
\end{figure}

We consider a UAV-assisted FEEL system comprising an edge server (ES) located at $S = (x_s, y_s, 0)$ and a set of UAVs denoted as $\mathcal{K} = \{1, 2, \dots, K\}$. The UAVs collaborate to train a computationally intensive global object detection model, specifically YOLO11x. Each UAV $k \in \mathcal{K}$ enters its designated non-overlapping rectangular mission area $\mathbb{A}_k$ at a starting point $p_k^{start}$ and finishes at an ending point $p_k^{end}$. As illustrated in Fig.~\ref{fig:model}, to rigorously capture the synchronization delay inherent in training such a heavy model, each global training round is modeled as a sequence of four phases: 1) downlink broadcasting, 2) distributed sensing, 3) local computation, and 4) model uploading. Let $T_k$ denote the total duration of these four phases for UAV $k$. Due to the synchronous protocol, the per-round completion time for the entire system is strictly bottlenecked by the slowest UAV, given by $T_{\max} = \max_{k \in \mathcal{K}} T_k$.

\subsection{Four-Phase  Round Structure}
\label{sec:structure}

\subsubsection{Downlink Broadcasting}
\label{sec:stage1}

At the beginning of the round, the ES broadcasts the global model (size $D_0$ bits) to all participating UAVs. Let $d_{b,k} = \Vert{}p_k^{start} - S\Vert{}$ denote the horizontal straight-line distance from the ES to the sensing starting point $p_k^{start} $. Suppose that  UAV $k$ flies   at a constant speed $v_{b,k} \in (0,  \mathrm{v}_{\max}]$  and altitude $h_k$, where $ \mathrm{v}_{\max}$ denotes the maximal allowable speed. During this flight, the horizontal distance $x_{b,k}$ between  UAV $k$ and the ES increases from 0 to $d_{b,k}$. Assuming a line-of-sight (LoS) dominated air-to-ground channel, the average downlink transmission rate $\bar{R}_{b,k}$ over this straight-line flight path is obtained as:
\begin{equation}
\bar{R}_{b,k} = \frac{1}{d_{b,k}} \int_{0}^{d_{b,k}} B_c \log_2\left(1 + \frac{\gamma_{b,k}}{(h_k^2 + x_{b,k}^2)^{\alpha/2}}\right) \mathrm{d}x_{b,k},
 \label{bkdy}
\end{equation}
where $B_c$ is the total channel bandwidth, $\gamma_{b,k}$  denotes the reference downlink  signal-to-noise ratio (SNR) received at UAV $k$, and $\alpha \ge 1$ is the path-loss exponent.
Since the UAV flies at a constant speed, the effective time-averaged transmission rate is   equivalent to the spatial average over the flight distance in (\ref{bkdy}).
To ensure the complete reception of the global model before the sensing task begins, UAV $k$'s received data volume within the broadcasting duration $T_{b,k} = d_{b,k}/v_{b,k}$ must  satisfy the following data volume constraint:
\begin{equation}
   T_{b,k} \cdot \bar{R}_{b,k} \ge D_0.
    \label{ICC:bD0int}
\end{equation}

\subsubsection{Distributed Sensing}
To acquire a spatially diverse training dataset in parallel, the global sensing task is distributed across $K$ UAVs. Upon reaching $p_k^{start}$, UAV $k$ executes a parallel-track (lawnmower) scanning trajectory over its designated area $\mathbb{A}_k$ (with length $L_k$ and width $W_k$). Note that  $p_k^{start}$ is heuristically chosen near the vertex of $\mathbb{A}_k$ closest to the ES to minimize the approach distance. The sensing flight is performed at a constant sensing speed $v_{s,k}$ and altitude $h_k$, subject to policy-mandated bounds:
\begin{equation}
    \mathrm{h}_{\min} \le h_k \le \mathrm{h}_{\max}, \quad \mathrm{v}_{\min} \le v_{s,k} \le \mathrm{v}_{\max},
    \label{P1:ganzhivh}
\end{equation}

A downward-facing camera is equipped at each UAV to capture  images at an interval of $\Delta t_{img}$. Given the horizontal and vertical fields of view (FOV) $\gamma_a$ and $\gamma_b$, the rectangular footprint of a single image has width $a_k = 2h_k \tan(\gamma_a/2)$ and length $b_k = 2h_k \tan(\gamma_b/2)$. To guarantee seamless coverage along the track, the displacement between consecutive images must not exceed the footprint length, which yields
\begin{equation}
v_{s,k} \Delta t_{img} \le b_k.
\label{P1:zhenjian}
\end{equation}
The lawnmower trajectory requires $N_{k,tr} = \lceil W_k / a_k \rceil$ parallel tracks to fully cover $\mathbb{A}_k$. Ignoring the negligible fractional track residuals for mathematical tractability, the total flight distance is tightly approximated by $L_{scan,k}(h_k) \approx N_{k,tr} L_k + W_k$. Consequently, the sensing duration is $T_{s,k} = L_{scan,k}(h_k) / v_{s,k}$, and  UAV $k$ concludes the sensing  at a determinable end point $p_k^{end}$ governed by the scanning geometry.

Since images are captured only along the tracks, the effective capture time is $T_{cap,k} = N_{k,tr} L_k / v_{s,k}$. This distributed sensing process forms a unique local dataset $D_k$  with size $|D_k| = \lfloor T_{cap,k} / \Delta t_{img} \rfloor$ for UAV $k$.
 It is noted that  increasing the altitude $h_k$ enlarges the footprint width $a_k$, which effectively reduces the required tracks $N_{k,tr}$ and shortens the sensing time $T_{s,k}$.  However, a higher altitude degrades the image spatial resolution, quantified by the GSD, defined as $\text{GSD}_k = 200 \tan(\gamma_a/2) h_k / r_a$, where $r_a$ is the pixel resolution. The sensing quality directly affects the YOLO11x detection confidence. Based on  empirical profiling \cite{drones9090595}, we model the successful participation probability of UAV $k$ as:
\begin{equation}
p_{s,k}(v_{s,k}, h_k) = \min(1, \max(0, C_k)),
\end{equation}
where $C_k$ is an empirical polynomial function of $v_{s,k}$ and $\text{GSD}_k$. Define a Bernoulli participation indicator $\mathbb{I}_k^{(n)}\! \in\! \{0, \!1\}$ with $\text{Pr}(\mathbb{I}_k^{(n)} \!= \!1) \!=\! p_{s,k}$. A successful sensing with $\mathbb{I}_k^{(n)} \!= \!1$ validates the dataset, allowing  UAV $k$ to proceed to the subsequent computation and uploading phases; otherwise, it aborts the task and returns to the ES at maximum speed.

\subsubsection{Local Computation}
 For a participating UAV with $\mathbb{I}_k^{(n)} = 1$, it hovers at the sensing end point $p_k^{end}$ to perform in-place model training. To flexibly trade off computation delay against learning performance, UAV $k$ selects a data subset of size $M_k \in \mathbb{Z}^+$ from $D_k$, and adapts its computation workload $M_k$ within the practical bounds
  \begin{equation}\label{P1:Mk}
   N_{\min,k} \le M_k \le N_{\max,k},
  \end{equation}
 where $N_{\max,k}=\lfloor |D_k|p_{s,k} \rfloor$ is the total acquired valid samples and $N_{\min,k}\ge 0$ guarantees a baseline training effectiveness.
 Let $w^{(n)}$ denote the global model parameters in the $n$-th round. The UAV computes the local updates via stochastic gradient descent (SGD) using its valid dataset  of size $M_k$. Let $\xi$ be the number of CPU cycles required to process one data sample, and $f_{cpu}$ be the CPU clock frequency. The local computation time is obtained as $T_{cp,k} = M_k \xi / f_{cpu}$. During this phase, the UAV consumes hovering propulsion power but provides a stable environment for computationally intensive training.

\subsubsection{Model Uploading} 
Upon completing the local training, the UAV flies back to the ES along the shortest straight line at a constant return speed $v_{u,k}$ to upload its model updates (assumed to be of size $D_0$).  Let $d_{up,k} = \Vert{}p_k^{end} - S\Vert{}$ denote the horizontal upload flight distance. To avoid mutual interference, frequency division multiple access (FDMA) is adopted, where UAV $k$ is allocated a bandwidth $B_k$, satisfying
\begin{equation}   \label{P1:daikuank}
  \sum_{k \in \mathcal{K}} B_k \le B_c, \quad B_k \geq 0,\ \forall k\in \mathcal{K}.
 \end{equation}
Similar to the broadcasting phase, the effective time-averaged uplink rate $\bar{R}_{u,k}$ during this straight-line return flight can be   expressed as a spatial average:
\begin{equation}\label{rup}
\bar{R}_{u,k} \!=\! \frac{1}{d_{up,k}} \int_{0}^{d_{up,k}} \!\!B_k \log_2\left(1 \!+\! \frac{\gamma_{u,k}}{(h_k^2\! +\! x_{u,k}^2)^{\alpha/2}}\right) \mathrm{d}x_{u,k},
\end{equation}
where $\gamma_{u,k}$ represents the reference uplink SNR received at the ES from UAV $k$ at the unit distance. The uploading duration $T_{up,k} = d_{up,k}/v_{u,k}$ must ensure the complete delivery of the updates, given by
\begin{equation}
T_{up,k} \cdot \bar{R}_{u,k} \ge D_0,\quad \forall k\in \mathcal{K}.
    \label{P1:uD0}
\end{equation}

 Finally, the ES aggregates the received updates from all participating UAVs to generate $w^{(n+1)}$. Given the abundant computational resources at the ES, the aggregation time cost   is negligible compared to the lengthy physical phases. Note that under the synchronous protocol, this aggregation step strictly awaits the arrival of the slowest UAV.

\subsection{Energy Consumption and Round Time}
\label{sec:energy}

In each  FEEL round, the total energy consumption of UAV $k$ is decoupled into operational energy for communication, sensing, and computation, and propulsion energy for flying and hovering. Let $P_r$, $P_s$, and $P_c$ denote the constant power for receiving, sensing, and uploading transmission at each UAV, respectively. The operational energy of a participating UAV is $E_{op,k} = P_r T_{b,k} + P_s T_{s,k} + \kappa f_{cpu}^3 T_{cp,k} + P_c T_{up,k}$, where $\kappa$ is the effective switched capacitance of the CPU. The propulsion power of a rotary-wing UAV at speed $v$ is modeled as \cite{li2020energy}:
\begin{equation}
P_{prop}(v) = P_0\Big(1+\frac{3v^2}{U_{tip}^2}\Big) + P_i \Big( \sqrt{1 + \frac{v^4}{4v_0^4}} - \frac{v^2}{2v_0^2} \Big)^{\frac{1}{2}} + \frac{1}{2}X v^3,
\end{equation}
where $P_0$, $P_i$, $U_{tip}$, $v_0$, and $X$ are hardware-specific aerodynamic constants. Since each phase involves either a constant-speed flight or a pure hover, the propulsion energy is evaluated in closed form as $E_{prop,k} = P_{prop}(v_{b,k})T_{b,k} + P_{prop}(v_{s,k})T_{s,k} + P_{prop}(0)T_{cp,k} + P_{prop}(v_{u,k})T_{up,k}$. For a dropout UAV with $\mathbb{I}_k^{(n)}=0$, it aborts the task and returns directly, thus consuming strictly less energy than a participant. To guarantee battery safety over a total of $N$ global training rounds, the worst-case energy must be bounded by the battery capacity $E_{k,\max}$:
\begin{equation} \label{P1:nenghao}
N \cdot (E_{op,k} + E_{prop,k}) \le E_{k,\max}, \quad \forall k \in \mathcal{K}.
\end{equation}
Under the synchronous FEEL protocol, the round time is bottlenecked by the slowest participant:
\begin{equation}\label{T(n)}
T^{(n)} = \max_{k \in \mathcal{K}} \left\{ T_{b,k} + T_{s,k} + \mathbb{I}_k^{(n)}\big(T_{cp,k} + T_{up,k}\big) \right\}.
\end{equation}
However, minimizing the stochastic maximum of expectations is mathematically intractable. To derive a trackable optimization problem, we introduce a continuous surrogate variable $T_{\max}$ to cap the expected completion time of every UAV. By replacing the discrete indicator $\mathbb{I}_k^{(n)}$ with its success probability $p_{s,k}$, the system bottleneck is effectively constrained by
\begin{equation}\label{P1:T_max}
T_{\max} \ge T_{b,k} + T_{s,k} + p_{s,k} (T_{cp,k} + T_{up,k}), \quad \forall k \in \mathcal{K}.
\end{equation}

\section{Convergence Analysis and Problem Formulation}\label{sec:conv}
\subsection{Convergence Analysis}\label{fenxi}

Following standard federated learning literature \cite{tang2025integrated}, we assume that the global loss function $F_S(w)$ is $L$-smooth and $\mu$-strongly convex, and let $w_*$ denote its optimal model parameters. We also assume unbiased local gradients with variance bounded by $\sigma_k^2/M_k$ and bounded data heterogeneity $\Lambda_k^2$ for all UAVs, since  $M_k$ data samples are computed.
\begin{theorem}
Assume that the learning rate satisfies $0 < \eta < 1/(4L)$. After $N > 0$ global rounds, the expected optimality gap of the global model is bounded by:
\begin{align} \label{slx:youbian}
\mathbb{E}[F_S(w^{(N)}) - F_S(w_*)] \le
\frac{G\left(1-A^N\right)}{\mu\eta(1-4L\eta)} + A^N \lambda,
\end{align}
where $A=1-\mu \eta(1-4L\eta)$, $\lambda = \mathbb{E}[F_S(w^{(0)})-F_S(w_*)]$ is the expected initial gap with $w^{(0)}$ being the initial model parameters, and $G$ is the steady-state error with the expression given in \cite[Theorem 1]{tang2025integrated}.
\end{theorem}
The proof closely follows  \cite{tang2025integrated} with minor algebraic adaptations for varying dataset sizes $M_k$, and is thus omitted due to space limits.
Moreover, existing works reveal that unequal participant probabilities $p_{s,k}(v_{s,k}, h_k)$ across devices severely inflate the steady-state error $G$ through data heterogeneity. To eliminate this statistical bias and ensure stable convergence, we mandate a uniform successful sensing probability for any arbitrary UAV $k \in \mathcal{K}$:
\begin{equation} \label{P1:xiangtong}
p_{s,k}(v_{s,k}, h_k) = p_s, \quad   p_{s,\min} \le p_s \le 1,
\end{equation}
where $p_{s,\min}$ is the predefined minimum reliability threshold.  Following (\ref{P1:xiangtong}), each UAV $k$ must adaptively adjust its sensing speed $v_{s,k}$ and altitude $h_k$ to achieve global unified  $p_s$. We further obtain Corollary \ref{cor:uniform}.
\begin{corollary}\label{cor:uniform}
 Under the uniform sensing probability $p_s$, an upper bound $\bar{G}$ to the steady-state error $G$ is found as 
\begin{align}
    G\le &\bar{G} = \frac{\eta}{K}\sum_{k\in \mathcal{K}}\Bigl(\frac{\sigma_k^2}{M_k}+\Lambda_k^2\Bigr) +\frac{4L\eta^2}{K}p_s^{-(K-2)}\sum_{k\in \mathcal{K}}\Lambda_k^2\nonumber \\
   &\quad +\frac{2L\eta^2}{K^2\chi p_s}\sum_{k \in \mathcal{K}}\Bigl(\frac{\sigma_k^2}{M_k}+2\Lambda_k^2\Bigr),
   \label{slx:shangjie}
\end{align}
 where $\chi = 1 - (1 - p_s)^K$.
 \end{corollary}
The proof to Corollary 1 is easy to obtain by applying the uniform $p_s$, and thus is omitted here for simplicity. Note that Corollary 1   reveals the intrinsic system tradeoffs: increasing the computation workload $M_k$ reduces the gradient variance $\sigma_k^2/M_k$ in $\bar{G}$, but prolongs the local computation time $T_{cp,k}$. Meanwhile, increasing $p_s$ suppresses the heterogeneity terms in $\bar{G}$, but forces the UAV's sensing parameters $(v_{s,k}, h_k)$ away from their time-optimal sensing settings. Hence, the variables $\{M_k, v_{s,k}, h_k\}$ must be jointly optimized. By replacing $G$ with $\bar{G}$, the RHS of
(\ref{slx:youbian}) is upper bounded by
\begin{equation}
\begin{aligned}
\Phi(\{M_k\},\{v_{s,k}\},\{h_k\},N)
=
\frac{\bar{G}(1-A^N)}{\mu\eta(1-4L\eta)}
+A^N\lambda. \notag
\end{aligned}
\end{equation}
Since $0<A<1$, the initial error term $A^N\lambda$
vanishes as $N$ increases, and $\Phi$ asymptotically
approaches $\bar{G}/[\mu\eta(1-4L\eta)]$.
To guarantee an $\epsilon$-optimal global model, where
$\epsilon>0$ is the predefined target optimality gap,
we impose the following convergence constraint:
\begin{equation}\label{P1:shangjieyueshu}
\Phi(\{M_k\},\{v_{s,k}\},\{h_k\},N)\leq\epsilon.
\end{equation}

\subsection{Problem Formulation}\label{sec:formulation}
We aim to minimize the expected total training time $N \cdot T_{\max}$, while guaranteeing the model convergence and meeting the multidimensional system limitations. Let $\boldsymbol{\Omega} \triangleq \big\{ \{v_{b,k}, v_{s,k}, v_{u,k}, h_k, M_k, B_k\}_{k\in\mathcal{K}}, p_s, T_{\max}, N \big\}$. The training time minimization problem is formulated as follows:
\begin{align}
    \text{(P1):} \quad \min_{\boldsymbol{\Omega}} \quad & N \cdot T_{\max} \nonumber \\
    \text{s.t.} \quad
        & (\ref{ICC:bD0int}) - (\ref{P1:zhenjian}),(\ref{P1:Mk}),(\ref{P1:daikuank}),(\ref{P1:uD0}), (\ref{P1:nenghao}), (\ref{P1:T_max}),(\ref{P1:xiangtong}), (\ref{P1:shangjieyueshu}), \nonumber\\
        & M_k \in \mathbb{Z}^+, \quad N \in \mathbb{Z}^+, \nonumber \\
        & v_{b,k}, v_{u,k} \in(0,\mathrm{v}_{\max}], \forall k \in \mathcal{K} \label{vmax}.
\end{align}
Due to the tight coupling of continuous  variables ($v_{s,k}, h_k$) and discrete  variables ($M_k, N$) in highly nonlinear constraints (e.g., (\ref{P1:nenghao}) and (\ref{P1:shangjieyueshu})), Problem (P1) is a mixed-integer nonconvex program (MINLP), which is generally NP-hard. Next section proposes an efficient alternating optimization (AO) framework to decouple and solve it.

\section{Two-Layer AO Algorithm}\label{sec:solution}

To solve (P1), we propose the Two-Layer AO algorithm. The outer layer conducts a  one-dimensional search over the rounds $N$ to strictly enforce the energy limit. 
For any given $N$, the inner AO iteratively updates four decoupled blocks: deriving $\{v_{b,k}\}$ in closed form, optimizing $\{v_{s,k}, h_k\}$ via successive convex approximation (SCA), determining $\{M_k\}$ through an integer-aware bisection search, and allocating $\{v_{u,k}, B_k\}$ in closed form.

\subsection{Optimal Broadcasting Speeds}\label{sec:sp1}

Given $\{M_k,v_{s,k},h_k,v_{u,k},B_k\}$,  (P1) reduces to the broadcasting subproblem (SP-1), which naturally decouples across the $K$ UAVs. For each UAV $k$, (SP-1) minimizes the broadcasting time $T_{b,k} = d_{b,k} / v_{b,k}$ over the speed $v_{b,k}$, subject to the data constraint in  (\ref{ICC:bD0int}) and the speed cap in (18).

Substituting  (\ref{bkdy}) into  (\ref{ICC:bD0int}) directly cancels the flight distance $d_{b,k}$, reformulating the data constraint in (\ref{ICC:bD0int}) as an explicit upper bound on the speed $v_{b,k} \le \frac{B_c I_k(h_k)}{D_0}$, where $I_k(h_k) \triangleq \int_0^{d_{b,k}} \log_2\left(1 + \frac{\gamma_{b,k}}{(h_k^2 + x_{b,k}^2)^{\alpha/2}}\right) dx_{b,k}$ is   independent of $v_{b,k}$. Because the objective $T_{b,k}$ monotonically decreases as $v_{b,k}$ increases, the optimal closed-form solution is directly obtained by capping this bound at the maximum speed:
\begin{equation}
v_{b,k}^* = \min\left\{ \mathrm{v}_{\max}, \frac{B_c I_k(h_k)}{D_0} \right\}.
 \label{ICC:vbstar}
\end{equation}

\subsection{Optimization of Sensing Phase}\label{sec:sp2}

Given $\{M_k, B_k, v_{b,k}, v_{u,k}\}$, (P1) reduces to the sensing subproblem (SP-2), which minimizes $T_{\max}$ over $\{v_{s,k}, h_k\}$ subject to  (\ref{P1:ganzhivh}),  (\ref{P1:zhenjian}),  (\ref{P1:T_max}),  (\ref{P1:xiangtong}), and  (\ref{P1:shangjieyueshu}). Problem (SP-2) is non-convex due to the fractional term $L_{scan,k}/v_{s,k}$ in (13), the nonlinear equality $p_{s,k}(v_{s,k}, h_k) = p_s$ in (15), the ceiling operation in the sensing distance, and the highly coupled variables in (17). The following addresses each constraint.

First, substituting the bound (16) into (17), the term $p_{s,k}^{-(K-2)}$ in (16) couples all UAVs. By replacing $p_{s,k}^{-(K-2)}$ with its worst-case constant $p_{s,\min}^{-(K-2)}$, we decouple the variables in (17), transforming it into a  lower bound on $p_s$:
\begin{equation}    \label{sjcx2}
p_s \ge \frac{\left[ \frac{2L\eta^2}{K^2 \chi_{\min}} \sum_{k\in\mathcal{K}} \left(\frac{\sigma_k^2}{M_k} + 2\Lambda_k^2\right) \right]}{A_0 - \frac{\eta}{K} \sum_{k\in\mathcal{K}} \left(\frac{\sigma_k^2}{M_k} + \Lambda_k^2\right) - \frac{4L\eta^2}{K p_{s,\min}^{K-2}} \sum_{k\in\mathcal{K}} \Lambda_k^2},
\end{equation}
where $A_0=\frac{(1-A)(\epsilon-\lambda A^N)}{1-A^N}$, $\chi_{\min}=1-(1-p_{s,\min})^K$.

Next, to handle  (13), we remove the ceiling operation  in $N_{k,tr}$ and newly obtain that $L_{scan,k} = W_k L_k / a_k + W_k$ with  sensing time $T_{s,k}(v_{s,k}, h_k) = L_{scan,k} / v_{s,k}$. Since both $T_{s,k}$ and $p_{s,k}(v_{s,k}, h_k)$ remain highly non-convex, directly solving (SP-2) is intractable. Hence, we employ SCA to replace these non-convex terms with their affine approximations. At the $i$-th iteration around a local point $(v_{s,k}^{(i)}, h_k^{(i)})$, their first-order Taylor expansions are derived as:
\begin{align}
\hat{T}_{s,k}^{(i)} &= T_{s,k}^{(i)} + \frac{\partial T_{s,k}}{\partial v_{s,k}}(v_{s,k} - v_{s,k}^{(i)}) + \frac{\partial T_{s,k}}{\partial h_k}(h_k - h_k^{(i)}),  \label{eq: SCA_T}\\
\hat{p}_{s,k}^{(i)} &= p_{s,k}^{(i)} + \frac{\partial p_{s,k}}{\partial v_{s,k}}(v_{s,k} - v_{s,k}^{(i)}) + \frac{\partial p_{s,k}}{\partial h_k}(h_k - h_k^{(i)}). \label{eq: SCA_p}
\end{align}
Substituting (\ref{eq: SCA_T}) and (\ref{eq: SCA_p}) into (13) successfully transforms the intractable expected time constraint into  a linear  inequality:
\begin{equation}  \label{TLZKTmax}
T_{b,k} + \hat{T}_{s,k}^{(i)} + \hat{p}_{s,k}^{(i)}(T_{cp,k} + T_{up,k}) \le T_{\max}, \quad \forall k \in \mathcal{K}.
\end{equation}
Similarly, the nonlinear coupling (15) is linearized as
\begin{equation}\label{TLZKTmax_p}
\hat{p}_{s,k}^{(i)} = p_s, \quad p_{s,\min} \le p_s \le 1.
\end{equation}

As a result, (SP-2) is approximated by the following convex problem at each iteration $i$:

$\begin{aligned}
\text{(SP-2.1):} \quad \min_{\makebox[0pt]{$\scriptstyle \{v_{s,k}\}, \{h_k\}, T_{\max}$}} \quad & \left[ T_{\max} + Z \right] \\  \text{s.t.} \quad & \text{(3), (4), (\ref{sjcx2}), (\ref{TLZKTmax})}, (\ref{TLZKTmax_p}), \\  & \vert{}v_{s,k} - v_{s,k}^{(i)}\vert{} \le \Delta_v^{(i)}, \quad\!\!\!\!\! \vert{}h_k - h_k^{(i)}\vert{} \le \Delta_h^{(i)},\end{aligned}$
where $Z = \frac{\rho}{2} \sum_{k\in\mathcal{K}} \left[ (v_{s,k} - v_{s,k}^{(i)})^2 + (h_k - h_k^{(i)})^2 \right]$ with $\rho > 0$ stabilizes the iteration, and the last two constraints define the trust region with step limits $\Delta_v^{(i)}$ and $\Delta_h^{(i)}$. At each iteration, problem (SP-2.1) can be efficiently solved by CVX to obtain the optimal variables, which then serve as the local points $(v_{s,k}^{(i+1)}, h_k^{(i+1)})$ for the next iteration. This SCA procedure repeats successively until the relative decrease of the objective $T_{\max}$ falls below a predefined tolerance $\delta$.

\subsection{Optimal Local Computation Workloads}\label{sec:sp3}

Given $\{v_{s,k},h_k,v_{b,k},B_k,v_{u,k}\}$, (P1) reduces to the workload subproblem (SP-3), which minimizes $T_{\max}$ over $\{M_k\}$ subject to (\ref{P1:Mk}), (\ref{P1:T_max}) and (\ref{P1:shangjieyueshu}). Since $T_{b,k}$, $T_{s,k}$, $T_{up,k}$ and $p_{s,k}$ are constants, the bottleneck time is determined by $T_{cp,k}$, which increases with $M_k$. Let $e_k=T_{b,k}+T_{s,k}+p_sT_{up,k}$ and $c=p_s\xi/f_{\mathrm{cpu}}>0$. The constraint in (\ref{P1:T_max}) is reduced to $e_k+cM_k\le T_{\max}$. For any candidate $T_{\max}=T$, the largest admissible integer workload for UAV $k$ is given by
\begin{equation}
\widehat{M}_k(T)=
\min\left\{N_{\max,k},
\left\lfloor\frac{T-e_k}{c}\right\rfloor\right\}.
\label{ICC:Mhat}
\end{equation}
A candidate $T$ is   feasible if $\widehat{M}_k(T)\ge N_{\min,k}$ for all $k$ and (\ref{P1:shangjieyueshu}) holds under $\{\widehat{M}_k(T)\}$. Since $\Phi$ in (\ref{P1:shangjieyueshu}) is nonincreasing over $M_k$, the feasibility of (SP-3) is monotonic with respect to $T$. To obtain the exact integer global optimum, we sort all possible time thresholds and construct the following set:
\begin{equation}
\mathcal{T}=\bigcup_{k\in\mathcal{K}}\left\{e_k+cm: m=N_{\min,k},\cdots,N_{\max,k}\right\},
\label{ICC:Tset}
\end{equation}
inside the operational range $[\max_k (e_k + c N_{\min,k}), \max_k (e_k + c N_{\max,k})]$. By performing a bisection search over the sorted set $\mathcal{T}$, the smallest feasible element    yields the optimal expected time $T_{\max}^*$ and the optimal workloads $M_k^* = \widehat{M}_k(T_{\max}^*)$.

\subsection{Optimal Uploading Speeds and Bandwidth}\label{sec:sp4}

Given $\{v_{b,k}, M_k, v_{s,k}, h_k\}$, (P1) reduces to the uploading subproblem (SP-4), which minimizes $T_{\max}$ over the uploading speeds $\{v_{u,k}\}$ and bandwidth allocations $\{B_k\}$, subject to (7), (9),  (13), and  (18). Similar to the broadcasting phase, since $T_{up,k} = d_{up,k}/v_{u,k}$, substituting the uploading rate (8) into (9) directly cancels $d_{up,k}$,  reformulating the data constraint as an upper bound on the uploading speed: $v_{u,k} \le \frac{B_k J_k(h_k)}{D_0}$, where $J_k(h_k) \triangleq \int^{d_{up,k}}_0 \log_2\left(1 + \frac{\gamma_{u,k}}{(h_k^2 + x_{u,k}^2)^{\alpha/2}}\right) dx_{u,k}$ is the spatial integral independent of $v_{u,k}$ and $B_k$.

 To minimize $T_{up,k}$, UAV $k$ must fly as fast as possible for any allocated bandwidth $B_k$. Thus, the time-optimal speed is tightly bounded by $v_{u,k} = \min\{\mathrm{v}_{\max}, \frac{B_k J_k(h_k)}{D_0}\}$. Let $T_{rem} = \max_{k\in \mathcal{K}}(T_{b,k}+T_{s,k}+p_sT_{cp,k})$ denote the fixed time budget consumed by the previous phases. According to the bottleneck constraint (13), the uploading time must satisfy $T_{up,k}\le (T_{\max}-T_{rem})/p_s$. Consequently, for a target maximum time $T$, enforcing $T_{up,k} \le (T-T_{rem})/p_s$ dictates the required speed $v_{u,k}(T) = p_sd_{up,k}/(T-T_{rem})$ and the required minimum bandwidth $B_k(T) = \frac{p_sD_0 d_{up,k}}{J_k(h_k)(T-T_{rem})}$. The target $T$ is globally feasible if and only if both $\sum_k B_k(T) \le B_c$ and $v_{u,k}(T) \le \mathrm{v}_{\max}$ hold. Since $\sum_k B_k(T)$ decreases with $T$, these two feasibility conditions give two  lower bounds on the time variable as $T \ge T_{bw}$ and $T \ge T_{cap}$, where $T_{bw}\triangleq T_{rem} + \frac{p_s D_0}{B_c}\sum_{k\in\mathcal{K}}\frac{d_{up,k}}{J_k(h_k)}$, $T_{cap}\triangleq\max_{k\in\mathcal{K}}\left(T_{rem} + p_s d_{up,k}/\mathrm{v}_{\max}\right)$.
Therefore, problem (SP-4) admits the closed-form optimal solutions: $T^{*}\!=\!\max\{T_{cap},T_{bw}\}$, $B_k^{*}\!=\!\frac{p_s D_0 d_{up,k}}{J_k(h_k)(T^{*}-T_{rem})}$, $v_{u,k}^{*}\!=\!\min\left\{\mathrm{v}_{\max}, \frac{p_s d_{up,k}}{T^{*}-T_{rem}}\right\}$.

\subsection{Overall Algorithm and Complexity}\label{sec:overall}

Algorithm 1 summarizes the proposed two-layer AO algorithm. The outer layer searches $N \in [N_{\min}^{r}, N_{\max}^{r}]$, respectively bounded by the convergence and energy constraints.
Since (SP-1), (SP-3), and (SP-4) are solved exactly and (SP-2) via SCA, the objective $T_{\max}$ is   non-increasing. Bounded below by zero, the inner loop  converges; and the global minimum is obtained by evaluating $T_{tot} = N \cdot T_{\max}$ across all feasible $N$.

Complexity: Let $I$ and $N_{\text{span}} = N_{\max}^{r} - N_{\min}^{r}$ denote the inner iterations and outer search size. Each iteration evaluates integrals in $\mathcal{O}(KQ)$, solves (SP-2.1) in $\mathcal{O}((2K+2)^4)$, and bisects $\mathcal{T}$ in $\mathcal{O}(K N_{\text{span}} \log(K N_{\text{span}}))$. Thus, the overall complexity is
 $\mathcal{O}\bigl(N_{\text{span}}\,I\,((2K+2)^4+KN_{\text{span}}\log(KN_{\text{span}})+KQ)\bigr)$,  which remains scalable for practical UAV swarms.

\begin{algorithm}[t]

\footnotesize
\setlength{\abovecaptionskip}{4pt}
\setlength{\belowcaptionskip}{2pt}

\caption{Two-layer AO Algorithm}
\label{alg:overall}
\begin{algorithmic}[1]
\Require $\mathcal{K}$, $S$, the sensing areas, $\epsilon$, $\delta$, and the system parameters.
\For{$N = N_{\min}^r$ to $N_{\max}^r$}
    \State Initialize the variables and set $i=0$;
    \Repeat
        \State Update $\{v_{b,k}\}$ and $\{B_k,v_{u,k}\}$ in closed form, $\{v_{s,k},h_k\}$ by (SP-2.1), and $\{M_k\}$ by (SP-3);
        \State $i=i+1$;
    \Until{the relative decrease of $T_{\max}^{(i)}(N)$ is below $\delta$}
    \If{$\Phi(\{M_k\},\{v_{s,k}\},\{h_k\},N)\le\epsilon$ and the energy budget holds for all $k$}
        \State Record $T_{\text{tot}}(N)=N\,T_{\max}^{(i)}(N)$ and the variables;
    \EndIf
\EndFor
\State $N^{*}=\arg\min_{N}T_{\text{tot}}(N)$.
\Ensure $N^{*}$ and the optimal $\{v_{b,k}^{*},v_{s,k}^{*},h_k^{*},M_k^{*},B_k^{*},v_{u,k}^{*}\}$.
\end{algorithmic}
\end{algorithm}

\section{Numerical Results and Performance Evaluation}\label{sec:sim}
\begin{figure}[t]
    \centering
    
    \includegraphics[width=0.94\columnwidth]{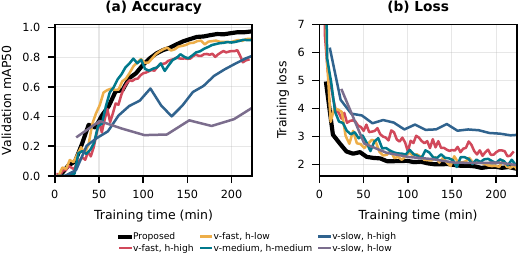}
    \caption{Comparison with Baseline~1 under fixed $(v_{s,k},h_k)$ from Table~\ref{tab:b1}: (a) validation mAP50 and (b) loss versus training time.}
    \label{fig:b1}
    
    \vspace{1.5em} 
    
    \includegraphics[width=0.94\columnwidth]{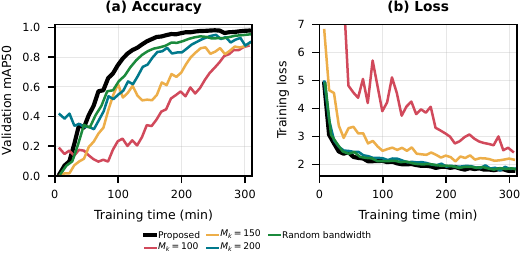}
    \caption{Comparisons with Baseline~2 under fixed $M_k\in\{100,150,200\}$ and Baseline~3 under a random bandwidth split: (a) validation mAP50 and (b) loss versus training time.}
    \label{fig:b2}
    
\end{figure}

\begin{table}[b]
\caption{System Parameters}
\label{tab:params}
\centering
\scriptsize
\renewcommand{\arraystretch}{0.78}
\setlength{\tabcolsep}{0pt}
\begin{tabular}{@{}l@{\hspace{4pt}}l@{\hspace{11pt}}l@{\hspace{4pt}}l@{}}
\toprule
Parameter & Value & Parameter & Value \\
\midrule
Number of UAVs $K$ & 2 & Model size $D_0$ & $217.61$ MB \\
Bandwidth $B_c$ & 6 MHz & SNR $\gamma_{b,k},\gamma_{u,k}$ & 52.5, 78.5 dB \\
Pixel array $r_a\times r_b$ & $4000\times3000$ & Diagonal FOV $\gamma_d$ & $84^{\circ}$ \\
CPU frequency $f_{cpu}$ & $5\times10^{8}$ Hz & Cycles per sample $\xi$ & $2.5\times10^{7}$ \\
Sensing speed $\mathrm{v}_{\min},\mathrm{v}_{\max}$ & 0.5, 30 m/s & Participation $p_{s,\min}$ & 0.84 \\
Altitude $\mathrm{h}_{\min},\mathrm{h}_{\max}$ & 8, 110 m & Energy budget $E_{k,\max}$ & 3.0 MJ \\
Image interval $\Delta t_{img}$ & 1 s & Path-loss exp. $\alpha$ & 2.3 \\
Powers $P_c,P_s,P_r$ & 5, 0.5, 0.5 W & Hardware constant $\kappa$ & $10^{-28}$ \\
\bottomrule
\end{tabular}
\end{table}
We evaluate the proposed system with two UAVs, each scanning a $200\times200$~m$^{2}$ area. For the federated learning task, the YOLO11x architecture is adopted and trained on the dataset provided in \cite{drones9090595}. The simulation parameters are listed in Table~\ref{tab:params}. The proposed joint optimization yields $N^* = 40$ FEEL rounds with an expected bottleneck time $T_{\max}^* = 458.96$s per round. The per-UAV total energy consumption reaches 2.98 MJ at  $N^* = 40$, rendering the 3.0 MJ battery constraint  in (\ref{P1:nenghao}) strictly tight. 
Fig. \ref{fig:b1} and Fig. \ref{fig:b2} compare our proposed algorithm against Baseline 1 with fixed sensing settings given in Table \ref{tab:b1} and Baselines 2 and 3 with fixed learning or resource allocations, respectively.

\subsubsection{Impact of  Sensing Parameters}
As compared to all the five settings in Baseline 1,  Fig.~\ref{fig:b1} shows that our  proposed algorithm always  achieves the fastest convergence and highest accuracy. Baseline 1 with fixed high-altitude settings, such as (v-slow, h-high), is observed to converge noticeably slower. This is because that although a higher altitude enlarges the camera footprint and reduces the required lawnmower tracks (thus shortening the sensing flight), it usually coarsens the  GSD  and degrades image quality. Hence, the   participation probability $p_{s}$ drops, demanding more time to converge. Conversely, the (v-slow, h-low) setting yields the worst mAP50. This is because that the low altitude shrinks the footprint, resulting in a   dense lawnmower trajectory. This inflates the sensing time, turning the UAV into a system straggler. Under a strict time budget, only a few aggregation rounds can be executed, severely limiting model generalization. Unlike Baseline 1, our proposed algorithm co-optimizes $\{v_{s,k}\}$ and $\{h_k\}$ along the $p_{s,\min}$ boundary, effectively balancing data quality against   time overhead.

\subsubsection{Impact of Data Volume and Resource Allocation}
While a large dataset  generally  improves per-round learning, training  YOLO11x may incur  large  in-place computation delay.
To balance this trade-off, our algorithm adaptively bounds the workloads, yielding optimal sizes $M_1^* = 120$ and $M_2^* = 119$ out of 227 and 228 captured samples.
Unlike our proposed algorithm, utilizing a fixed data size in Baseline 2 fails to balance this trade-off. Specifically, as shown in Fig. 3, 
a large dataset with $M_k = 200$ severely prolongs the round time, causing a rightward convergence shift, whereas a small dataset with $M_k = 100$ fundamentally limits the final accuracy. Moreover, Baseline 3’s random bandwidth allocation preserves per-round accuracy but induces unoptimized uploading delays. This causes a system straggler that visibly shifts its training curve rightward. In contrast, our closed-form allocation optimally equalizes uploading times across UAVs, thereby eliminating this straggler effect.

\begin{table}[!t]
\caption{Baseline 1: UAV Sensing Speed and Height Intervals}
\label{tab:b1}
\centering
\scriptsize
\renewcommand{\arraystretch}{0.84}
\setlength{\tabcolsep}{2.5pt}
\begin{tabular}{lccc}
\toprule
Scenario & Speed (m/s) & Altitude (m) & Fixed $(v_s,h)$ \\
\midrule
v-fast, h-high & $[7,10.69]$ & $[70,110.37]$ & $(9,90)$ \\
v-fast, h-low & $[7,10.69]$ & $[8,45]$ & $(9,30)$ \\
v-medium, h-medium & $[5,8]$ & $[45,70]$ & $(5.5,55)$ \\
v-slow, h-high & $[0.5,4]$ & $[70,110.37]$ & $(1,90)$ \\
v-slow, h-low & $[0.5,4]$ & $[8,45]$ & $(1,25)$ \\
\bottomrule
\end{tabular}
\end{table}

\section{Conclusion}
We investigated UAV-assisted federated edge vision. We conducted convergence analysis and  enforced a uniform participation probability to mitigate sensing data heterogeneity. To minimize the total training time under per-UAV operation and energy constraints, we proposed the Two-layer AO algorithm to jointly optimize the training rounds, sensing mobility, computation workloads, and bandwidth allocations. Simulations validate that our design significantly accelerates model convergence and accuracy.
Future work will relax the linear flight assumption, optimizing UAV trajectories during broadcast and upload phases for time minimization.
 
\bibliographystyle{IEEEtran}
\bibliography{IEEE}

\end{document}